\documentstyle[psfig]{article}
\begin{document}

\centerline{\Large \bf Spectral properties of the X-ray binary pulsar}
\centerline{\Large \bf LMC~X$-$4 during different intensity states}
\bigskip
\centerline{\bf S. Naik and B. Paul}
\centerline {Tata Institute of Fundamental Research, Mumbai, 400 005, India.}

\bigskip
\section*{Abstract}
We present spectral variations of the binary X-ray pulsar LMC~X$-$4 
observed with the RXTE/PCA during different phases of its 30.5 day long third
period. Only out of eclipse data were used for this study. 
The 3$-$25 keV spectrum, modeled with high energy cut-off power-law and
iron line emission is found to show strong dependence on the intensity state.  
Correlations between the Fe line emission flux and different parameters of 
the continuum are presented here.

\medskip
{\it Key-words}{ . stars: individual: LMC~X$-$4 -- stars: neutron -- X-rays: stars}

\section{Introduction}
LMC~X$-$4 is an eclipsing high-mass disk-fed accretion-powered binary 
X-ray pulsar in the Large Magellanic Cloud. A spin period of 13.5 s was
discovered in LMC~X$-$4 by Kelley et al. (1983) and X-ray eclipses with 
a 1.4 day recurring period was discovered by Li et al. (1978) and White 
(1978). The X-ray intensity varies by a factor of $\sim$ 60 between high 
and low states with a periodic cycle time of 30.5 day (Lang et al. 1981; 
Paul \& Kitamoto 2002). Flux modulation at super-orbital period in LMC~X$-$4 
is believed to be due to blockage of the direct X-ray beam by its precessing 
tilted accretion disk, as in the archetypal system Her~X$-$1. Flaring events
of duration ranging from $\sim$ 20 s to 45 minutes (Levine et al. 1991 and
references therein) are seen about once in a day during which the source
intensity increases by factors up to $\sim$ 20.

Broad band spectroscopy using GINGA and ROSAT data shows that the continuum
can be modeled with a high energy cut-off power-law (Woo et al. 1996). 
The spectrum also shows a soft excess and a broad iron emission line. The soft
excess detected with ROSAT was modeled as combination of thermal bremsstrahlung
and very soft black-body by Woo et al. (1996), while the same observed with
Beppo-SAX (La Barbera et al. 2001) was modeled as black-body emission from 
accretion disk at magnetoshperic radius Comptonized by moderately hot electrons.
La Barbera et al. (2001) also reported the presence of a cyclotron absorption 
line at $\sim$ 100 keV.

In this paper we present the spectral variations of LMC~X$-$4 during the
30.5 day long period using the archival data from RXTE observations.

\begin{figure}[h]
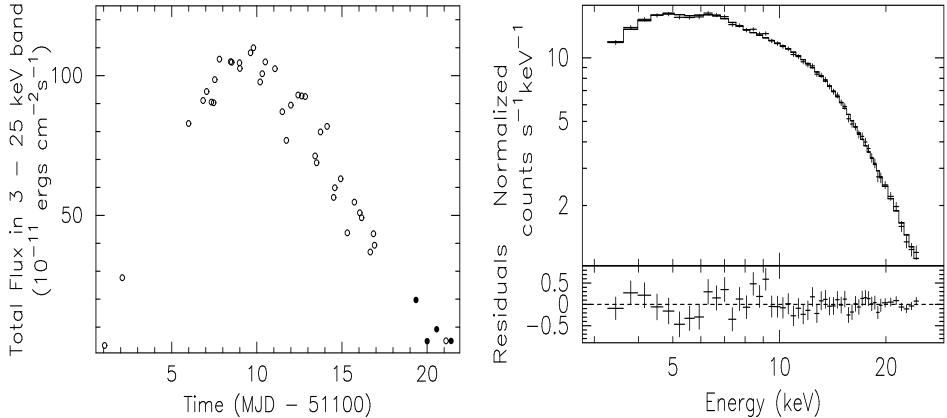

\begin{center}
\hbox{
\psfig{figure=mjd.ps,height=5.5cm,width=6.0cm,angle=-90}
\hspace{0.2cm}
\psfig{figure=sp.ps,height=5.5cm,width=6.0cm,angle=-90}
}
\end{center}
\vspace{-0.9cm}
\caption{\footnotesize
{{\it Left panel}: Average background-subtracted X-ray flux in 3 $-$ 25 keV 
energy range obtained from the RXTE/PCA observations of LMC~X$-$4 in 1998. The
points marked by ``$\bullet$'' are  for the observations which were made
outside the selected time range and have been included here based on the phase
of the super-orbital period. These observations are used to get a better 
coverage of low intensity state.
{\it Right panel}: The figure shows the observed count rate spectrum of 
LMC~X$-$4 on 1998 October 22. The best fit model consists of a blackbody 
(kT $=$ 0.2 keV), a power law and a high energy cutoff. The iron emission line 
was kept fixed at 6.4 keV with width of 0.65 keV. }}
\label{fig1}
\end{figure}

\section{Observation, Analysis and Results}

To study the super-orbital phase dependence of various spectral parameters, 
we have analyzed 43 RXTE/PCA observations of LMC~X$-$4 at different phases 
of the 30.5 day third period. The data used for analysis are out-of-eclipse and 
free from the the flaring state. Energy spectra in 129 channel were generated
from the Standard 2 mode PCA data. The standard procedures for data selection,
and response matrix generation were followed. Background estimation was done
using both bright and faint models of RXTE/PCA according to different 
intensity states of the source at different phases. We restricted our analysis 
to 3 $-$ 25 keV energy range. Data from all five PCUs are added together. We 
have fitted the energy spectrum of the source using a model consisting of 
blackbody, power law and a high energy cutoff as model components. We have 
included a Gaussian line near the expected K$_\alpha$ emission from iron and 
absorption edge due to iron. The value of equivalent hydrogen column density 
N$_H$ was set to have a lower threshold of 0.055 $\times$ 10$^{22}$ cm$^{-2}$ 
which is the Galactic column density towards this source. The blackbody
temperature was kept fixed at kT $=$ 0.2 keV, while the center and width of
the iron emission line was fixed at 6.4 keV and 0.65 keV respectively with
free normalization. The variation in 3 $-$ 25 keV source flux during the
RXTE/PCA observations are shown in the left panel of Figure 1 whereas the
right panel of Figure 1 shows the energy spectrum for one of the 
observations at high intensity state. The variation of iron line flux and 
iron equivalent width with the source flux in 7$-$25 keV energy range are 
shown in the left and right panels of Figure 2 respectively.

\begin{figure}[h]
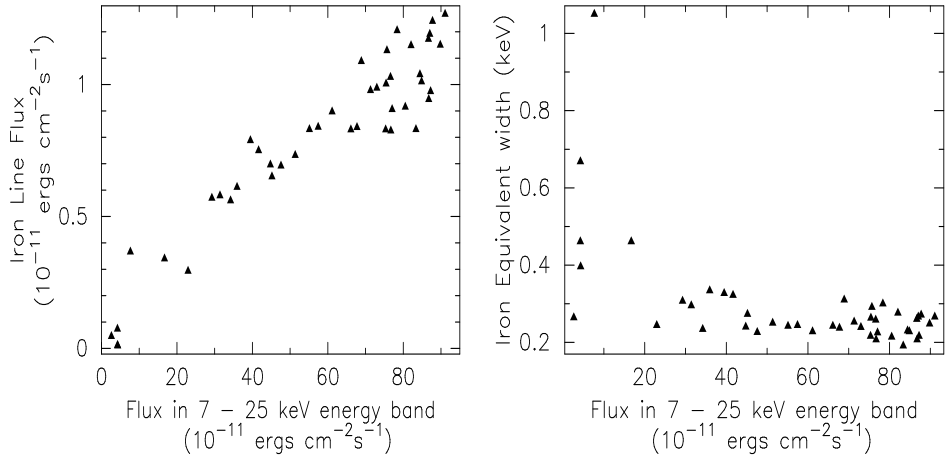

\begin{center}
\hbox{
\psfig{figure=lmc_f1.ps,height=6.0cm,width=6.0cm,angle=-90}
\hspace{0.2cm}
\psfig{figure=lmc_f2.ps,height=6.0cm,width=6.0cm,angle=-90}
}
\end{center}
\vspace{-0.9cm}
\caption{\footnotesize
{{\it Left panel}:The figure shows the variation in iron line flux with the
source flux in 7 $-$ 25 keV energy band. 
{\it Right panel}: The variation in equivalent width of the iron emission line
with the source flux in 7 $-$ 25 keV energy range is shown. It is observed
that the iron equivalent width is high when the source flux is low.}}
\label{fig2}
\end{figure}

The results obtained from this work are summarized as follows.

(i) The iron emission line flux is found to be directly correlated with
the source flux in 7 $-$ 25 keV energy range.

(ii) The source spectrum is found to be flat with power-law photon index 
in the range 0.5 $-$ 0.7 during low intensity state (source flux $\leq$ 3 
$\times$ 10$^{-10}$ ergs cm$^{-2}$ s$^{-1}$ at 3$-$25 keV energy range) whereas
during high intensity state, the spectrum is steep with power-law photon index
in the range 0.7 $-$ 0.9.

(iii) Equivalent width of the iron emission line is found to be highly 
variable in the range 0.25 $-$ 1.1 keV during low intensity state (source 
flux $\leq$ 2 $\times$ 10$^{-10}$ ergs cm$^{-2}$ s$^{-1}$ in 7$-$25 keV 
energy range), whereas it remains almost constant (0.2 $-$ 0.35 keV) during 
the high intensity state with source flux $\geq$ 2 $\times$ 10$^{-10}$ ergs 
cm$^{-2}$ s$^{-1}$ in 7$-$25 keV energy range.

\section{Discussion}

According to the results of the present work, it is observed that iron 
intensity correlates very well with the continuum intensity in 7 $-$ 25 
keV energy range.  The equivalent width of the iron emission line is very 
high at low luminosity. Similar thing was found in Vela X$-$1 (Becker et al.
1979; White et al. 1983). In Vela X$-$1, it can be due to increase in 
absorption caused by the stellar wind of the primary. 

Nagase et al. (1986) studied the change in equivalent widths of iron emission
line against the column density of matter in the line of sight for Vela~X$-$1.
Similar studies were done for GX~301$-$2 by Makino et al. (1985) and for 
Her~X$-$1 by Makishima (1986). These results suggest that the column density 
averaged over the whole direction does not change appreciably, whereas the 
absorption column density along the line of sight changes drastically with 
time and orbital phase. Inoue (1985) and Makishima (1986) estimated the 
equivalent widths of the fluorescence iron line emission from neutral matter 
in a sphere surrounding the X-ray source using a power law type incident 
spectrum. They found that, if the matter is located between the X-ray source 
and the observer, the continuum spectrum is absorbed by the matter resulting in 
increasing the equivalent width monotonically with the column density as 
observed in GX~301$-$2 (Makino et al. 1985). However, when the X-ray source 
is hidden by some thick material, the equivalent width remains almost constant 
($\sim$ 1 keV) (as observed from Vela~X$-$1 during the eclipse; Nagase et al. 
1984). In case of accretion powered X-ray pulsars, if the compact object is 
hidden from direct view by the accretion disk and only X-rays scattered into 
the line of sight by an accretion disk corona or wind are visible, the iron 
equivalent width can be higher. This may explain the higher value of iron 
equivalent width during the low intensity states of LMC~X$-$4.

\section{Acknowledgment}
Work of SN is partially supported by the Kanwal Rekhi Scholarship of
the TIFR Endowment Fund. This research has made use of data obtained 
through the High Energy Astrophysics Science Archive Research Center 
Online Service, provided by the NASA/Goddard Space Flight Center.

{}
\end{document}